\documentstyle[psfig,twocolumn,floats,prl,aps]{revtex}
\input epsf.tex

\begin{document}
%\draft

\title{Tree structure of the percolating Universe}
\author{S. Colombi$^1$, D. Pogosyan$^2$ and T. Souradeep$^3$ }
\address{1. Institut d'Astrophysique de Paris, CNRS, 98 bis boulevard
Arago,
F-75014 Paris, France}
\address{2. CITA, 60 Saint George street, Toronto, Ontario M5S 3H8,
Canada}
\address{3. Department of Physics, Kansas State University, Manhattan,
Kansas 66506-2601}
\date{\today}
\maketitle

\begin{abstract}
We present a numerical study of topological descriptors of initially
Gaussian and scale-free density perturbations evolving via
gravitational instability in an expanding universe. We carefully
evaluate and avoid numerical contamination in making accurate
measurements on simulated fields on a grid in a finite
box. Independent of extent of non linearity, the measured Euler number
of the excursion set at the percolation threshold, $\delta_{\sc c}$,
is positive and nearly equal to the number of isolated components,
suggesting that these structures are trees.  Our study of critical
point counts reconciles the clumpy appearance of the density field at
$\delta_{\sc c}$ with measured filamentary local curvature. In the
Gaussian limit, we measure $|\delta_{\sc c}|> \sigma$ in contrast to
widely held belief that $|\delta_{\sc c}| \sim \sigma$, where $\sigma^2$
is the variance of the density field.
\end{abstract}


% body of paper here

%%\section{Introduction}
The large scale structure of the universe can be characterized
topologically at various scales by studying the excursion sets of the
density contrast $\delta ({\bf x})$; i.e., the sets $E^+_{\delta_{\sc
th}}=\{{\bf x} | \delta({\bf x}) \ge \delta_{\sc th}\}$ or
$E^-_{\delta_{\sc th}}=\{{\bf x} | \delta({\bf x}) \le \delta_{\sc
th}\}$. An interesting choice of $\delta_{\sc th}$ in either cases is
the percolation threshold, $\delta_{\sc c}$, for which the excursion
set includes an infinite, connected structure~\cite{zel82}.

Accurate determination of $\delta_{\sc c}$ is a challenge both from
the theoretical point of view -- no analytic calculation of
$\delta_{\sc c}$ exist so far in the three-dimensional case; the
experimental point of view -- numerous spurious effects, such as
discreteness, finiteness of the sampled volume, etc., can contaminate
the measurements, and the results depend on the method
employed~\cite{bar83,kly,dom92}.  Here, we proceed as in
Ref.~\cite{dom92} and consider site percolation on a grid, i.e.,
neighboring sites above/below the density threshold are connected
through faces.

We analyze the topology of $E^\pm_{\delta_{\sc th}}$ in terms of the
three dimensional local curvature of $\delta({\bf x})$ given by the
Hessian matrix, $\partial^2 \delta/\partial x_i \partial x_j$. The
number of negative eigenvalues, $I$, of the Hessian matrix at any
point classifies the local density structure into one of four distinct
classes: region with $I=3$ is in a ``clump''(c); $I=2$ in a
``filament'' (f); $I=1$ in a ``pancake'' (p) and $I=0$ in a
``void''(v). In particular, this classification applies to critical
points of the field, where the gradient $\partial\delta/ \partial
x_i=0$. Connectivity of the excursion set $E^\pm_{\delta_{\sc th}}$ is
dictated solely by the number, ${\cal C}_I$, of critical points of
each class $I$ within $E^\pm_{\delta_{\sc th}}$. For clarity, we
replace the numeral index $I$ of ${\cal C}_I$ by the alphabetic label
$I=0,1,2,3 \rightarrow {\rm v},{\rm p}, {\rm f}, {\rm c}$.

At a simple qualitative level, the role of critical points in
outlining the connectivity of excursion sets can be understood quite
intuitively. Connectivity at percolation is along special field lines
(``ridges'' and ``river beds'') threading the critical
points. Filament saddle points (f-saddles) are thus crucial for
percolation in overdense regions ($E^+_{\delta_{\sc th}}$).  Indeed, a
large fraction of them lie along ridges connecting two local maxima.
When $\delta_{\sc th} > \delta $ at a saddle point, the point drops
out of the excursion set, thus disconnecting the corresponding clumps.
Analogous reasoning can be applied to percolation in underdense
regions where a large fraction of pancake saddle points (p-saddles)
lie on river beds that connect local minima.

%%\vspace{-10.5cm}
%%\hfill\hfill {\bf KSUPT-00/1 }
%%\vspace{10.cm}

In detail, the situation can be more complicated than this simple
vision, e.g., saddle-points connected through ridges or river
beds. However, the work of Morse rigorously establishes a simple link
between the number distribution of critical points (local property
related to curvature) and global connectivity (topology) of the
excursion sets of any smooth field lending credence to our above
intuitive line of argument~\cite{morse}.  A simple, but powerful,
relation exists between the Euler characteristic, $\chi$~\cite{dor},
of $E^\pm_{\delta_{\sc th}}$ and a linear combination of the four
critical point counts, ${\cal C}_I$.  It can indeed be shown that
despite all the possible complexity~\cite{morse},
\begin{equation}
\chi^\pm = \pm \left[\ {\cal
           C}_{\rm c}-{\cal C}_{\rm f}+{\cal C}_{\rm p} -{\cal C}_{\rm
           v} \right].
\label{morseeqn}
\end{equation}
%We define a `reduced' Euler number, ${\cal E}^\pm\equiv\chi^\pm/{\cal
%C}_{\rm tot}$, where ${\cal C}_{\rm tot}=\sum_I {\cal C}_I$ is the
%total number of critical points.

The observed structures in the galaxy distribution are believed to
have evolved through gravitational instability from small initial
fluctuations, usually postulated to be a Gaussian random field. We pay
particular attention to the Gaussian limit since at large scales the
memory of initial conditions is retained in the evolved system.  At
small scales, non linear clustering creates an asymmetry between
overdense and underdense regions possibly changing the properties of
the percolating network and the value of $\delta_{\sc c}$. To study
clustering in the non linear regime we use $N$-body simulations.

In this paper we have two goals: (i) obtaining an accurate measurement
of $\delta_{\sc c}$ as a function of smoothing scale; (ii)
characterizing the topology of the percolating network through a
combination of the Euler number, critical points and local curvature.
We attempt to exploit eq.~(\ref{morseeqn}) to understand the global
properties, such as percolation, of the evolving density contrast in
the universe in terms of the distribution of (local) critical
points. 

%Since eq.~(\ref{morseeqn}) applies to any continuous and
%smooth field, one attractive feature of our approach is that the
%density fields in the linear as well as the non-linear regimes can be
%studied on a common basis. In particular, we find that results in the
%linear Gaussian regime do carry over into the non linear regime.
 
In our numerical experiment, we consider $N$-body simulations
(described in~\cite{hivon}) with scale-free Gaussian initial
conditions, where the power-spectrum as a function of wavenumber, $k$,
spectral index, $n$, and amplitude, $A$, is given by $P(k)= A k^n$,
$n=-2,-1,0$. This covers the observationally relevant range of values
in the scaling regime $1 \lesssim \ell \lesssim 100$ Mpc.  The
simulations were generated with a particle-mesh code~\cite{pm} with
$256^3$ particles in a periodic cubic box of size $L$. A mesh
resolution $N_{\rm r}=256$ was used to compute the forces. The density
field was computed using Cloud-In-Cell interpolation~\cite{he} on
grids of various sizes $N_{\rm g}=64,128,256$ and then convolved with
Gaussian windows of scales $\ell/L=1/N_{\rm g}$, $2/N_{\rm g},\ldots,
\,0.04$, $0.08$ to insure sufficient
differentiability~\cite{footnote}. The corresponding variance of the
smoothed density distribution will be denoted by $\sigma^2(\ell)$.  To
increase the dynamic range of our measurements, we take advantage of
self-similar behavior for scale-free initial conditions~\cite{Pee} and
combine several time snapshots of each simulation with the appropriate
scaling $(L,\ell)\rightarrow (L/s, \ell/s)$, where $s$ is the linear
theory correlation length ($\sigma^2_{\rm linear}(s)\equiv 1$). For
Gaussian smoothing, $s=[a^2 A\ \Gamma[(n+3)/2]/(4\pi^2)]^{1/(n+3)}$,
where $a$ is the expansion factor of the Universe, arbitrarily fixed
to unity at the beginning of the simulation.

To study the Gaussian limit, a set of $10$ pure Gaussian realizations
of the density field, $\delta({\bf x})$, at each grid size and
spectral index, $(N_{\rm g},n)$, were generated and smoothed at the
same set of scales, $\ell/L$, as in the $N$-body samples.

The gradient and the Hessian of the density field are computed by
least square fitting a second order hypersurface
to the values $\delta$ computed at each site and its 26 closest
neighbors. 

We have explored the wide range of smoothing scales $\ell/L$ and
box sizes $N_{\rm g}$ to thoroughly quantify the numerical effects
contaminating topological properties. The optimal set of
parameters $(N_{\rm g},\ell/L)=(256,0.01)$ is adopted to minimize 
the following contamination effects (partly discussed in
\cite{dom92}):

(i) The grid must be sufficiently thin compared to the smoothing
scale, $N_{\rm g}\ell/L \gtrsim 2.5$. With this constraint,
our procedure for computing numerical derivatives of $\delta$ provides
extremely accurate results.

(ii) Periodic boundaries enhance percolation in a finite box but the
effect is negligible if $\ell/L \ll 1$. On the other hand, finite box
size suppresses large scale $[{\cal O}(L)]$ power, which can be viewed
as an increased effective spectral index.  From our Gaussian samples,
we find that combination of these two competing effects is
inconsequential if $\ell/L \lesssim 0.01$.

(iii) Inherent limitations of the $N$-body approach: spatial
resolution of the forces imposes $N_{\rm r} \ell/L \gtrsim 1$. To
evade effects of $N$-body relaxation and unrealistic non linear
couplings between the large ${\cal O}(L)$ modes, it is required that
$2/N_{\rm g}^{1/3} \lesssim s/L$ and $s/L \lesssim 0.1$, respectively.
%Possible transients due to the fact the initial configuration
%is realized using the Zel'dovich approximation~\cite{sco} are 
%neglected.
Finally, discreteness effects are insignificant since an $8$-fold
dilution of the number of particles does not change the results
substantially.

\begin{figure*}[tb]
\centerline{\psfig{figure=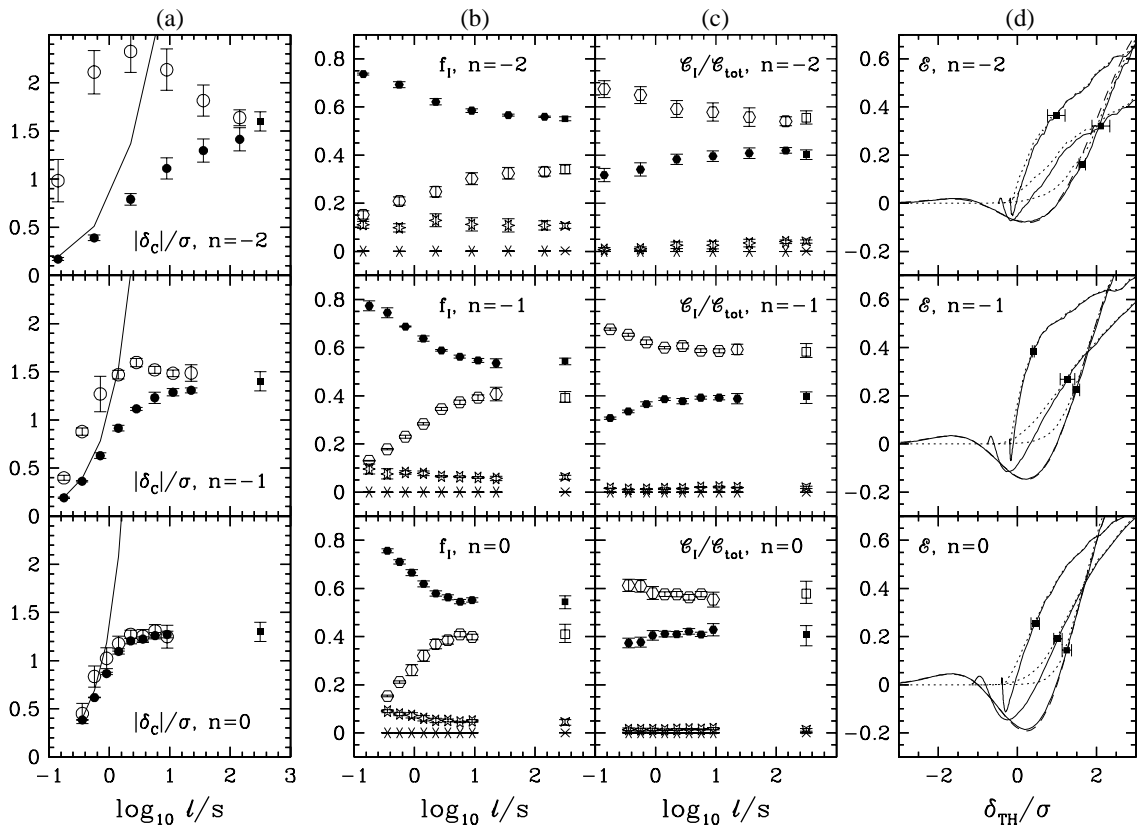,bbllx=55pt,bblly=341pt,bburx=541pt,bbury=696pt,width=15cm}}
\caption[]{Topological properties of scale-free universes with
Gaussian initial conditions. (a) The measured percolation threshold
$\delta_{\sc c}$ as a function of scale. Each panel corresponds to the
spectral index $n$ indicated on it, respectively $n=-2$, $-1$, $0$
from top to bottom. Open (filled) circles correspond to measurements
in the $N$-body simulations for overdense (underdense) regions. The
errorbars represent deviations from self-similarity. The rightmost
square gives the Gaussian limit obtained from the average over $10$
random realizations. The solid curve is $1/\sigma(\ell/s)$,
corresponding to the expected asymptotic regime $\delta_{\sc c}^{-}
\rightarrow -1$ for percolation of underdense regions in the limit
$\ell/s\rightarrow 0$.  (b) The measured fractions, $f_I$, of
clump/filament/pancake/void regions in $E^+_{\delta_{\sc c}}$ as
functions of scale (respectively, open and filled circles, stars and
asterisks). Again, the rightmost points give the Gaussian limit, where
the $f_I$'s can also be computed analytically as functions of
$\delta_{\sc th}$~\cite{usagain}.  (c) As in (b) but the fractions of
the four types of critical points, ${\cal C}_I/{\cal C}_{\rm
tot}$. (d) The measured Euler characteristic as a function of density
threshold at three different stages of the simulations, as explained
in the text (solid curves). The number of isolated clusters $N_{\rm
clus}/{\cal C}_{\rm tot}$ is shown by dots. The solid curve that
corresponds to initial conditions matches perfectly the analytic
Gaussian limit~\cite{dor} (dashes). Symbols with horizontal error bar
indicate the measured value of the percolation threshold, $\delta_{\rm
c}$.}
\label{fig1}
\end{figure*}

First column of panels in Fig.~\ref{fig1} shows the measured
percolation threshold, $\delta_{\sc c}$, as a function of scale,
$\ell/s$. In agreement with previous works~\cite{yes,sah}, the
asymmetry due to gravitational clustering between overdense ($+$) and
underdense regions ($-$), $\delta_{\sc c}^{+} > |\delta_{\sc c}^{-}|$,
increases with the level of non linearity and with $-n$. It is very
small for $n=0$. For underdense regions, the asymptotic regime
$\delta_{\sc c}^{-} \rightarrow -1$ is naturally reached when $\ell/s
\ll 1$.  The value of $\delta_{\sc c}^{\rm +}/\sigma$ is maximum near
transition to the nonlinear regime (see also Ref.~\cite{kly}), $\ell/s
\sim 1$, where initial structures are optimally enhanced by
large-scale motions~\cite{web}, and is larger than the Gaussian limit,
the difference increasing with $-n$. As noted later, $\delta_{\sc
c}^{+}/\sigma$ can become small in the highly nonlinear regime due to
lost coherence.
 
In the Gaussian limit, we find slight increase in $\delta_{\sc c}$
with $-n$; $|\delta_{\sc c}/\sigma|=1.6 {\pm0.1}$, $1.4 {\pm0.1}$,
$1.3 {\pm0.1}$ for $n=-2$,$-1$,$0$, respectively. Error bars reflect
the dispersion in the $10$ realizations and possible contaminations
(i) and (ii) discussed above. These results should be contrasted with
that of Ref.~\cite{yes}, where $|\delta_{\sc c}/\sigma| \simeq 1$ is
obtained independently of $n$. However, the measurements in
Ref.~\cite{yes} use a smaller grid, $N_{\rm g}=64$, and might be
affected by effects (ii) mentioned above.  The other difference is
that the field is softened by a power-spectrum cut-off at small scales
instead of our Gaussian smoothing.

Second column of panels in Fig.~\ref{fig1} displays the fraction $f_I$
of space occupied by clump, filament, pancake and void regions in
$E^+_{\delta_{\sc c}}$. As shown earlier from a qualitative point of
view~\cite{sah2}, the simulated universes at percolation are seen to
be composed mostly of filamentary and clumpy regions, in the
respective fractions of $~55 \%$ and $~35-40\%$ in the Gaussian limit.
The filamentary nature becomes dominant in the non linear regime and
the results do not depend significantly on $n$.

\begin{figure}[tbh]
\centerline{\psfig{figure=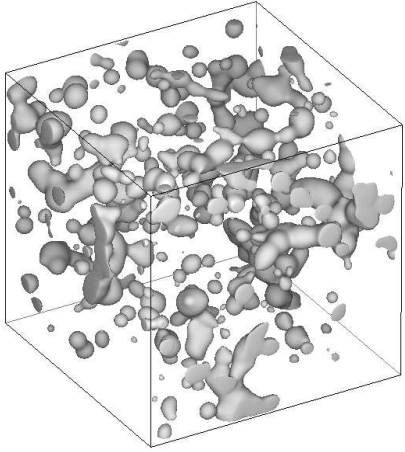,bbllx=81pt,bblly=146pt,bburx=528pt,bbury=645pt,width=5.5cm}}
\caption[]{3-D view of the overdense percolating network in the $n=-1$
simulation, at $a=128$ and smoothed at $\ell=0.02$.  It consists of
spherical-like high density regions dominated by 'clump' points
($I=3$) connected by bridges built mainly from points with filamentary
local curvature ($I=2$). No significant number of closed overdense
loops is present.  }
\label{fig2}
\end{figure}

Third column of panels in Fig.~\ref{fig1} shows the measured fraction
of critical points of each kind. Note that maxima are dominant,
explaining the rather clumpy appearance of the largely filamentary (in
terms of curvature) percolating network in Fig.~\ref{fig2}. In fact,
clumpiness augments with non linearity as the fraction of maxima
increases and explains the rather small $\delta_{\sc c}^{+} /\sigma$
despite increased filamentarity when $\sigma \gg 1$.

Fourth column of panels in Fig.~\ref{fig1} shows the ``reduced'' Euler
number ${\cal E} \equiv \chi / {\cal C}_{\rm tot}, ~ {\cal C}_{\rm
tot}=\sum_I {\cal C}_I$, as a function of density threshold measured
at various stages of the simulations, corresponding to $a=1,32,128$
for $n=0,-1$ and $a=1,16,32$ for $n=-2$. The value of $a$ increases
with the abscissa of the left local maximum of ${\cal E}$. For $a=1$,
the measurements agree almost perfectly with the analytic prediction
in the Gaussian limit~\cite{dor}, as expected. The ${\cal E}$ curves
retain overall shape with time, except that they are progressively
compressed near the origin and stretched at large $\delta_{\sc
th}/\sigma$. This perhaps indicates that non linear clustering does
not significantly influence the topology of the system.

There are three regimes dictated by the sign of the first derivative
of ${\cal E}$. With increasing $\delta_{\sc th}/\sigma$, (i) at small
$\delta_{\sc th}/\sigma$, mostly local minima drop out of
$E^+_{\delta_{\sc th}}$, creating cavities, thus increasing ${\cal
E}$; (ii) at intermediate $\delta_{\sc th}/\sigma$, p-saddles start to
drop out too, and cavities connect together, thus the value of ${\cal
E}$ decreases; (iii) in the large $\delta_{\sc th}/\sigma$ region even
f-saddles drop out, breaking up the ridges to create isolated
clusters, thus ${\cal E}$ increases again. At this last stage, the
contribution of minima and p-saddles to the Euler number can be
largely neglected, $\chi \approx {\cal C}_{\rm c}-{\cal C}_{\rm f}$. 

Fourth column of Fig.~1 embodies the main result of the paper.  It
shows that percolation occurs at positive values of the Euler
characteristic $\chi > 0$ in regime (iii), and specifically at $\chi
\simeq N_{\rm clus}$, where $N_{\rm clus}$ is the number of ``isolated
clusters'' (disconnected regions of $E^+_{\delta_{\sc th}}$). This
means that at percolation threshold, isolated clusters, including the
largest percolating one, have close to tree structure\cite{tree},
i.e.~they contain no or few loops.  Indeed, in this case each cluster
contributes unity to the total Euler number adding up to $\chi \approx
N_{\rm clus}$.  The f-saddles, the only type available in number in
the regime (iii) can produce only loops, which decrease Euler number
of individual clusters, and being abundant would decrease the total
$\chi$ much below $ N_{\rm clus}$, which is not what is observed. Thus
we conclude that the percolating network is a tree.
Fig.~\ref{fig2} gives a visual presentation of a percolating tree.

%Results for underdense regions are largely analogous to the overdense
%case~\cite{usagain}. However, there are a few significant
%differences. In the non linear regime, void fraction is reduced in
%favor of filaments rather than pancakes, although the latter remain
%dominant.

In summary, we used very simple but novel statistics in cosmology --
{\em critical points of the density field and its local curvature} to
characterize the topological properties of the universe at
percolation.  The most important results of our
analyses~\cite{usagain} are the following:

(a) At percolation, the Euler characteristic is {\em positive}, equal
to the number of isolated clusters suggesting that most of these have
{\em tree} structure (with almost no loops). This result is valid for
overdense as well as underdense excursions.  In particular, we find in
the Gaussian limit that $|\delta_{\sc c}|/\sigma >1 $, at variance
with common belief (e.g. Ref.~\cite{yes}) that site percolation in
$E^\pm_{\delta_{\sc th}}$ should occur at $|\delta_{\sc c}|/\sigma =
1$, where the Euler number $\chi^\pm$ of $E^\pm_{\delta_{\sc c}}$ is
zero.

(b) In terms of curvature, the overdense structures at percolation are
mostly filamentary and clumpy in agreement with previous
works~\cite{yes,web,sah2}.  Filamentarity increases with level of non
linearity. However, maxima are the dominant critical points in the
excursion set. With isotropic smoothing, this explains its rather
clumpy appearance as illustrated by Fig.~\ref{fig2}.

%(d) The value of $\delta_{\sc c}/\sigma$ for percolation
%of overdense regions is maximum around the 
%transition to the non linear regime and
%larger than the Gaussian limit by a difference that increases with
%$-n$.

Our efforts to identify and address sources of contamination in our
numerical experiments suggest that it is actually difficult to measure
the percolation threshold accurately in a real galaxy catalog, where
one of the main issues is contamination by discreteness effects. These
impose a large smoothing scale, a fact difficult to
reconcile with the requirement $\ell/L \lesssim 0.01$, even for
current or forthcoming large redshift surveys such as the Two-degree
Field~\cite{2dF} or the Sloan Digital Sky Survey~\cite{SDSS}.

However, it is still possible to measure local curvature and critical
point counts, because these statistics should not require as small
values of $\ell/L$ to be measured with sufficient accuracy. Although,
here we focus on critical point counts at the percolation threshold,
these counts as functions of $\delta_{\rm TH}$ are of interest. In
particular, the statistics ${\cal C}_I$ and $f_I$ can be computed
explicitly as functions of $\delta_{\rm TH}$ in the Gaussian
limit~\cite{usagain}. As the genus of the boundary of
$E^\pm_{\delta_{\sc th}}$~\cite{gottetal}, they could be used in real
galaxy catalogs to constrain the nature of the primordial density
field. To do that properly, several issues remain to be addressed,
such as the anisotropies induced by redshift distorsion in
three-dimensional galaxy catalogs or the effects of biasing between
the observed galaxy distribution and the underlying total matter
distribution.

We thank E. Hivon for kindly providing the $N$-body simulations and
S. F. Shandarin and J. R. Bond for fruitful discussions. TS
acknowledges support through NSF CAREER grant AST-9875031. Computer
time for the $N$-body simulations was allocated by the scientific
council of IDRIS, Orsay. DP \& TS thank IAP and SC thanks CITA for
hospitality.
% now the references. delete or change fake bibitem. delete next three
%   lines and directly read in your .bbl file if you use bibtex.


\begin{references}
\bibitem{footnote}  Gaussian smoothing at scale $\ell$ is
roughly comparable to averaging in a sphere of radius $\ell \sqrt{5}$.
\bibitem{tree} Tree refers to a simply connected set (or graph),i.e.,
all paths between any two points can be smoothly deformed into
each other.
\end{references}
\end{document}